\documentstyle[procl,epsfig]{article}

\def\be{\begin{equation}}
\def\ee{\end{equation}}
\def\bea{\begin{eqnarray}}
\def\eea{\end{eqnarray}}

\newcommand{\bq}{\begin{equation}}
\newcommand{\eq}{\end{equation}}

\newcommand{\ba}{\begin{eqnarray}}
\newcommand{\ea}{\end{eqnarray}}

\newcommand{\lesim}{\raisebox{-0.05cm}{$\stackrel{<}{{\scriptstyle
 \sim}}$} }

\begin{document}

\begin{titlepage}

\begin{flushleft}
DESY 96--012 \\
{\tt hep-ph/9601352} \\
July 1996 (revised)
\end{flushleft}

\setcounter{page}{0}

\mbox{}
\vspace*{\fill}
\begin{center}
{\Large\bf Constraining the Proton's Gluon Density}\\

\vspace{1.8mm}
{\Large \bf $\!\!\!$by Inclusive Charm Electroproduction at 
HERA$^{\ast}\!\!\!$}\\

\vspace{5em}
\large
Andreas Vogt

\vspace{2em}
\normalsize
{\it Institut f\"ur Theoretische Physik, Universit\"at W\"urzburg}\\
{\it Am Hubland, D--97074 W\"urzburg, Germany} \\
\normalsize
\vspace*{\fill}
{\large \bf Abstract}
\end{center}
\vspace*{3mm}
We analyze the capability of charm production in deep-inelastic $ep$
scattering at HERA to constrain the gluon distribution $g(y,\mu ^2)$
of the proton.  The dependence of the theoretical predictions for the
charm structure function $F_{2}^{\, c}$ on the mass factorization
scale $\mu $ and the charm mass is investigated. $F_{2}^{\, c}$ seems
to be well suited for a rather clean and local gluon measurement at
small momentum fractions $y$.

\noindent

\vspace*{\fill}
$^{\ast}$ Invited talk presented at the {\it International Workshop on
Deep Inelastic Scattering and Related Phenomena (DIS'96)}, Rome, Italy, 
April 15--19, 1996. To appear in the proceedings.


\end{titlepage}

\title{
 CONSTRAINING THE PROTON'S GLUON DENSITY
 BY INCLUSIVE CHARM ELECTROPRODUCTION AT HERA}

\author{ ANDREAS VOGT }

\address{
 Institut f\"ur Theoretische Physik, Universit\"at W\"urzburg, \\
 Am Hubland, D--97074 W\"urzburg, Germany}

\maketitle\abstracts{
We analyze the capability of charm production in deep-inelastic $ep$
scattering at HERA to constrain the gluon distribution $g(y,\mu ^2)$
of the proton.  The dependence of the theoretical predictions for the 
charm structure function $F_{2}^{\, c}$ on the mass factorization 
scale $\mu $ and the charm mass is investigated. $F_{2}^{\, c}$ seems 
to be well suited for a rather clean and local gluon measurement at 
small momentum fractions $y$.}

\noindent
Recent measurements of the structure function $F_{2}^{\, ep}$ at HERA
rather directly determine the (light) quark densities in the proton at 
very small momentum fractions $y$. On the other hand, extractions of the 
gluon distribution from $F_{2}$ scaling violations are more indirect 
and require assumptions on the evolution kernels at small $y$, e.g.\ of 
the validity of the usual NLO evolution equations, which themselves are 
a matter of investigation presently. 
An obvious candidate for a more direct gluon determination is the 
longitudinal structure function $F_L$, but in practice this quantity 
is not easily measured sufficiently accurate over a wide kinematical 
range. In this context it is interesting to recall that $F_2$ itself is 
expected to contain an appreciable part directly sensitive to the gluon 
density at small Bjorken-$x$, namely its charm production contribution 
$F_{2}^{\, c} $~\cite {LRSN1}. In the following we discuss some 
theoretical aspects of this observable relevant to the determination of 
$g(y,\mu^2)$; for the experimental status at HERA see ref.~\cite{Daum}.

In next-to-leading order (NLO) perturbative QCD, the electromagnetic
(exchange of one photon with virtuality $Q^2$) structure function 
$F_{2}^{\, c}$ reads~\cite{LRSN}
\small
\begin{eqnarray}
  F_{2}^{\, c}(x,Q^2) & \!\!\! = \!\!\! & 
    \frac{Q^2 \alpha_S(\mu ^2)}{4 \pi^2 m_{c}^{2}} \int_{ax}^{1} \! 
    \frac{dy}{y} \, yg(y,\mu^2) \, e_{c}^{2} \left\{ c_{2,g}^{\, (0)} 
    + 4 \pi \alpha_S(\mu ^2) \left[ c_{2,g}^{\, (1)} + \bar{c}_{2,g}
    ^{\, (1)} \ln \frac{\mu^2}{m_{c}^{2}} \right] \right\}\! \nonumber \\
                      &   & \!\!\!\!\!\!\!\!
    \mbox{} \:\: + \frac{Q^2 \alpha_{S}^{2}(\mu ^2)}{\pi m_c^{2}} \sum_
    {q=u,d,s} \int_{ax}^{1}\! \frac{dy}{y} \, y(q+\bar{q})(y,\mu^2)
    \left[ c_{2,q}^{\, (1)} + \bar{c}_{2,q}^{\, (1)} \ln \frac{\mu^2}
    {m_{c}^{2}} \right] \:\:\: . 
\end{eqnarray}
\normalsize
Here $m_c$ ($e_c$) denotes the charm mass (charge), and $\alpha_S$
is the strong coupling constant. The renormalization scale has been put 
equal to the ($\overline{\mbox{MS}}$) mass factorization scale $\mu $. 
The lower limit of integration over the fractional initial-parton
momentum $y$ is given by $ y_{\rm min} = ax = (1 + 4m_{c}^{2}/Q^2) x $, 
corresponding to the threshold $\hat{s} = 4m_{c}^{2} $ of the partonic 
center-of-mass (c.m.) energy squared. 
The NLO coefficient functions $c^{(1)}$ and $\bar{c}^{(1)}$ have been 
calculated in ref.~\cite{LRSN}, and a convenient parametrization of 
these results has been provided \cite{RSN} in terms of 

\small
\begin{equation}
  \xi  = \frac{Q^2}{m_{c}^{2}} \:\:\: , \:\:\:\:\: 
  \eta = \frac{\hat{s}}{4m_{c}^{2}} - 1 
       = \frac{\xi}{4} \left( \frac{y}{x} -1 \right) -1 \:\:\: .
\end{equation}
\normalsize
In leading order (LO), ${\cal O}(\alpha_S)$, $F_{2}^{\, c}$ is sensitive 
only to $g(y,\mu^2)$ via the well-known Bethe-Heitler process 
$ \gamma^{\ast}g \rightarrow c\bar{c}$ \cite{BH}. A comparison of the 
various contributions to $F_{2}^{\, c}$ in NLO shows that for the 
physically reasonable scales $\mu$, $ \mu \simeq 2 m_c \ldots \sqrt{Q^2 
+ 4 m_{c}^{2}} $ (see below), the quark contribution in (1) --- which 
is not necessarily positive due to mass factorization --- is very small,
about 5\% or less. Therefore $F_{2}^{\, c}$ does represent a clean 
gluonic observable also in NLO. 

\begin{figure}[htb]
\vspace{-4mm}
\begin{center}
\mbox{\epsfig{file=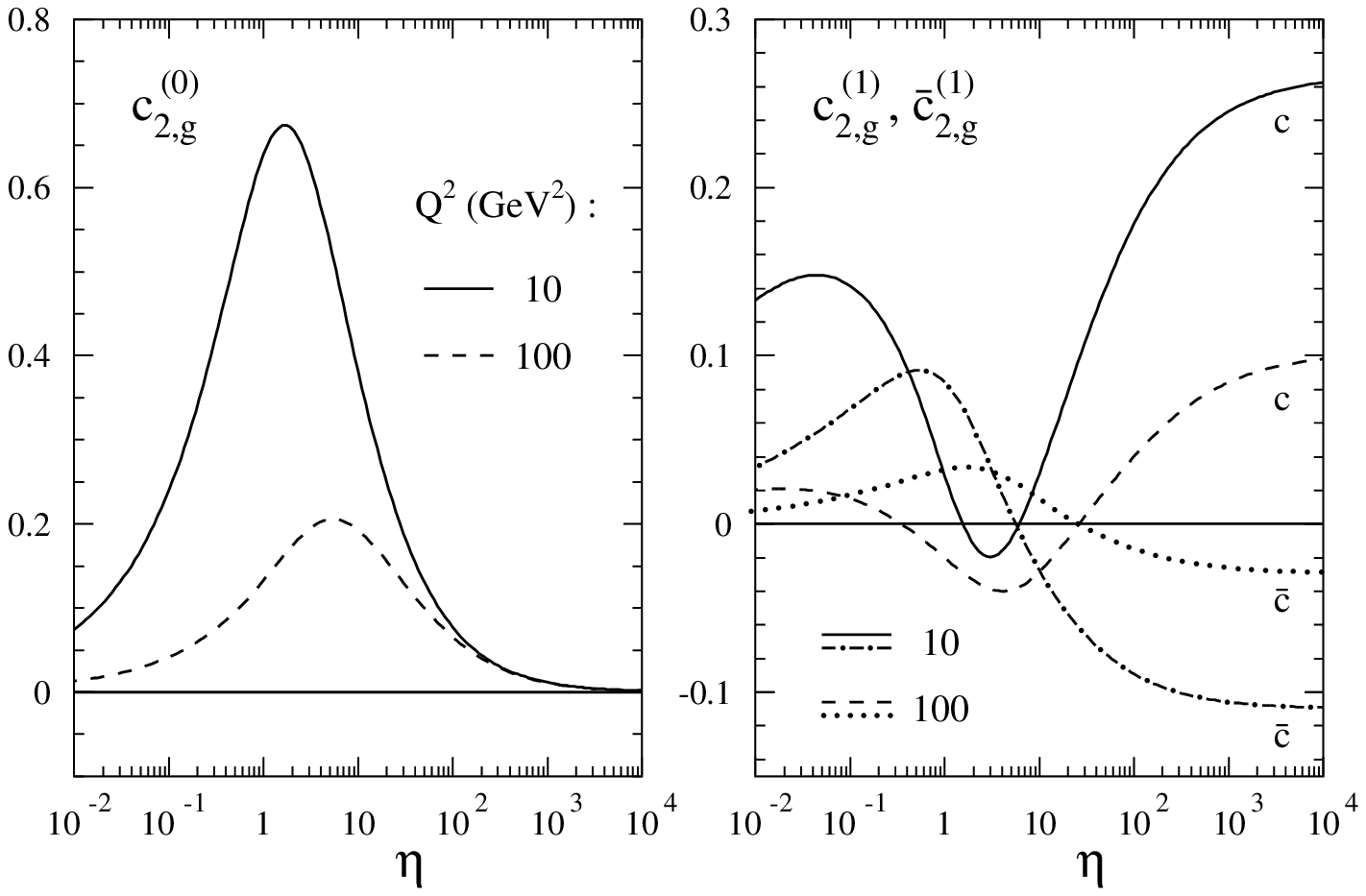,height=4.95cm,width=9.8cm}}
\vspace*{-3mm}
\end{center}
{\footnotesize \sf 
  Figure 1: The $\eta $-dependence of the coefficient functions
  $c_{2,g}^{\, (0,1)}$, $\bar{c}_{2,g}^{\, (1)}$ derived in refs.$^{\, 
  3-5} $ for two values of $Q^2$ with $m_c\! =\! 1.5$ GeV. The scheme-%
  dependent quantity $c_{2,g}^ {\, (1)}$ is given in the $\overline
  {\mbox{MS}}$ scheme.}
\vspace{-5mm}
\end{figure}
The $\eta$-dependence of the gluonic coefficient functions$^{\, 3-5}$ is 
recalled in Fig.~1 for two values of $Q^2$ typical for deep-inelastic 
small-$x$ studies at HERA, as in the following using $ m_c =1.5 $ GeV. 
The comparison of the NLO coefficients $c^{(1)}$ and $\bar{c}^{(1)}$
with the LO result $c^{(0)}$ reveals that potentially large corrections 
arise from regions where $c^{(0)}$ is small, i.e.\ from very small and 
large partonic c.m.\ energies.
These corrections are due to initial-state-gluon bremsstrahlung and the
Coulomb singularity at small $\hat{s}$, and due to the flavour 
excitation (FE) process at $\hat{s} \gg 4m_{c}^{2} $ ($\eta \gg 1 $). 
For a more detailed discussion, including the quark contributions, see 
ref.~\cite{LRSN}. Large FE-logarithms have been resummed --- at the 
expense of losing the full small-$\hat{s}$ information of (1) --- by 
introducing a charm parton density, leading to the so-called variable-%
flavour scheme \cite{VFS}. For another approach to high $\hat{s}$ see 
ref.~\cite{CCH}. The importance of these corrections in the HERA 
small-$x$ regime considered here will be investigated below.

The first question to be addressed in order to judge the phenomenolo%
gical usefulness of $F_{2}^{\, c}$ as a gluon constraint is whether or
not the available NLO expression (1) is sufficient for obtaining results
which are stable under variation of the (unphysical) mass factorization 
scale. It has been argued~\cite{GRS} that one should use $\mu \simeq 
2m_c $, since $\mu $ is supposed to be controlled by $\hat{s}$ and the 
integrand in (1) is maximal close to the lower limit, $ \hat{s} \simeq 4 
m_{c}^{2} $. The range of significant contributions in $\hat{s}$, 
however, broadens considerably with increasing $Q^2$ (see below), hence 
$ \mu \simeq \sqrt{Q^2 + 4m_{c}^{2}} $, chosen in 
refs.~\cite{LRSN1,LRSN,RSN}, appears at least equally reasonable. 
Therefore we estimate the theoretical uncertainty of $F_{2}^{\, c}$ in 
NLO by varying the scale between $\mu = (2)m_c$ and $ \mu = 2 
\sqrt{Q^2 + 4m_{c}^{2}} $, see Fig.~2.
\begin{figure}[htb]
\vspace{-4mm}
\begin{center}
\mbox{\epsfig{file=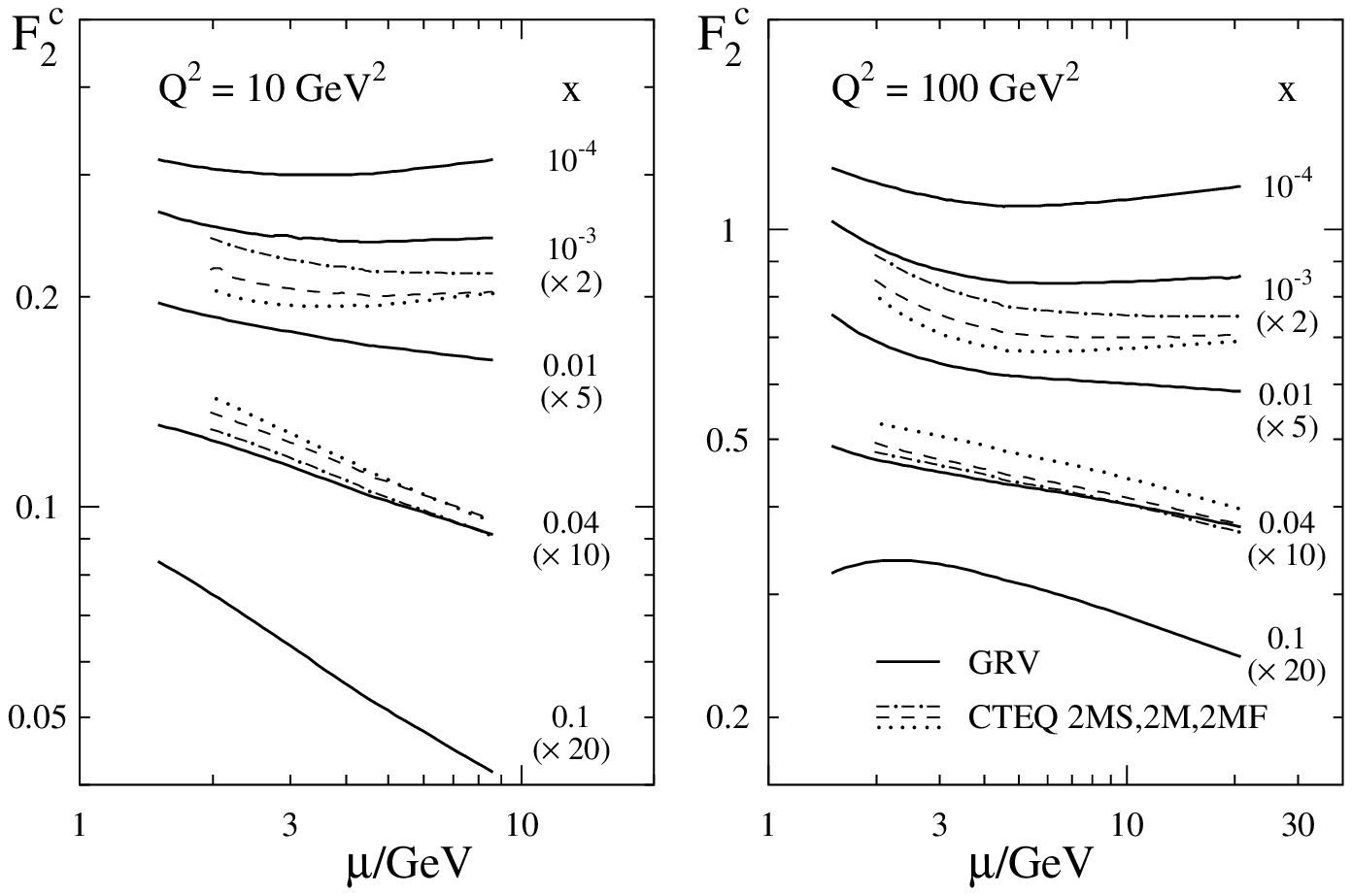,height=4.95cm,width=9.8cm}}
\vspace*{-3mm}
\end{center}
{\footnotesize \sf
  Figure 2: The dependence of $F_{2}^{\, c}$ on the scale $\mu $ for
  $ m_c \leq \mu \leq 2 \sqrt{Q^2 + 4 m_{c}^{2}} $ at selected
  values of $x$ and $Q^2$. GRV~\cite{GRV94} and CTEQ2~\cite{CTEQ} 
  (at $ x = 10^{-3}, \, 0.04 $) parton densities have been employed.}
\vspace{-2mm}
\end{figure}

\noindent
One finds that at small-$x$, $x\, \lesim \, 10^{-2}$, the scale variation
amounts to at most about $\pm 10\% $. Moreover, the scale stability at 
small $x$ does not significantly depend on the steepness of the gluon 
distribution.
Consequently, the NLO results of ref.~\cite{LRSN} seem to provide rather
sound a theoretical foundation for a small-$x$ gluon determination at 
HERA, despite the large total c.m.\ energy $ s \gg 4m_{c}^{2} $ which 
might suggest a destabilizing importance of $ \ln [\hat{s}/(4m_{c}^{2})]$
terms.
At large $x$, $ x \approx 0.1 $, on the other hand, the scale dependence
of $F_{2}^{\, c}$ is rather strong, especially at low $Q^2$, presumably
due to the large small-$ \eta $ threshold contributions mentioned
above.  However, $F_{2}^{\, c}$ is small in this region.

The next issue we investigate is the locality in $y$ of the gluon 
determination via $F_{2}^{\, c}$. The contribution from initial-parton 
momenta smaller than $y_{\rm max}$ to $F_{2}^{\, c}(x,Q^2)$, denoted by 
$F_{2} ^{\, c}(x,Q^2, y_{\rm max}) $, is presented in Fig.~3 for the GRV 
parton distributions \cite{GRV94}. At scales $ \mu \approx \sqrt{Q^2 + 
4m_{c} ^{2}}$, about 80\% of $F_{2}^{\, c}$ originates in the region 
$ y_{\rm min}\! =\! ax \leq y \:\lesim\: 3 y_{\rm min} $. Again the 
situation is very similar for the CTEQ2 parton densities \cite{CTEQ}. 
Thus $F_{2} ^{\, c}$ allows for rather local a determination of 
$g(y,\mu^2 \approx \sqrt{Q^2 + 4m_{c} ^{2}})$. 
The partonic c.m.\ energies in the region contributing 80\% to the 
structure function $F_{2}^{\, c}$ are given by $\eta \, \lesim \, 5$ 
(20), corresponding to $\hat{s} \, \lesim \, 60\, (180) $ GeV$^2$, at 
$ Q^2 = 10\, (100) $ GeV$^2$, respectively. This implies that for the 
$Q^2$ values under consideration here, the plateau region of $c
^{\, (1)}$ and $\bar{c}^{\, (1)}$ at large $\eta $ (c.f.\ Fig.~1)
does not play an important role. Resummation of large-$\eta $ FE 
logarithms is thus not necessary, and not appropriate if it implies 
additional approximations in the more important small-$\hat{s}$
region~\cite{VFS}. A similar observation has already been made in 
\cite{GRS} for $\mu \simeq 2m_{c}$ at low $Q^2$.  The latter scale 
choice leads however to a considerably wider important range 
of $\hat{s}$ at high $Q^2$, somewhat in contrast to its motivation 
described above.
\begin{figure}[htb]
\vspace{-4mm}
\begin{center}
\mbox{\epsfig{file=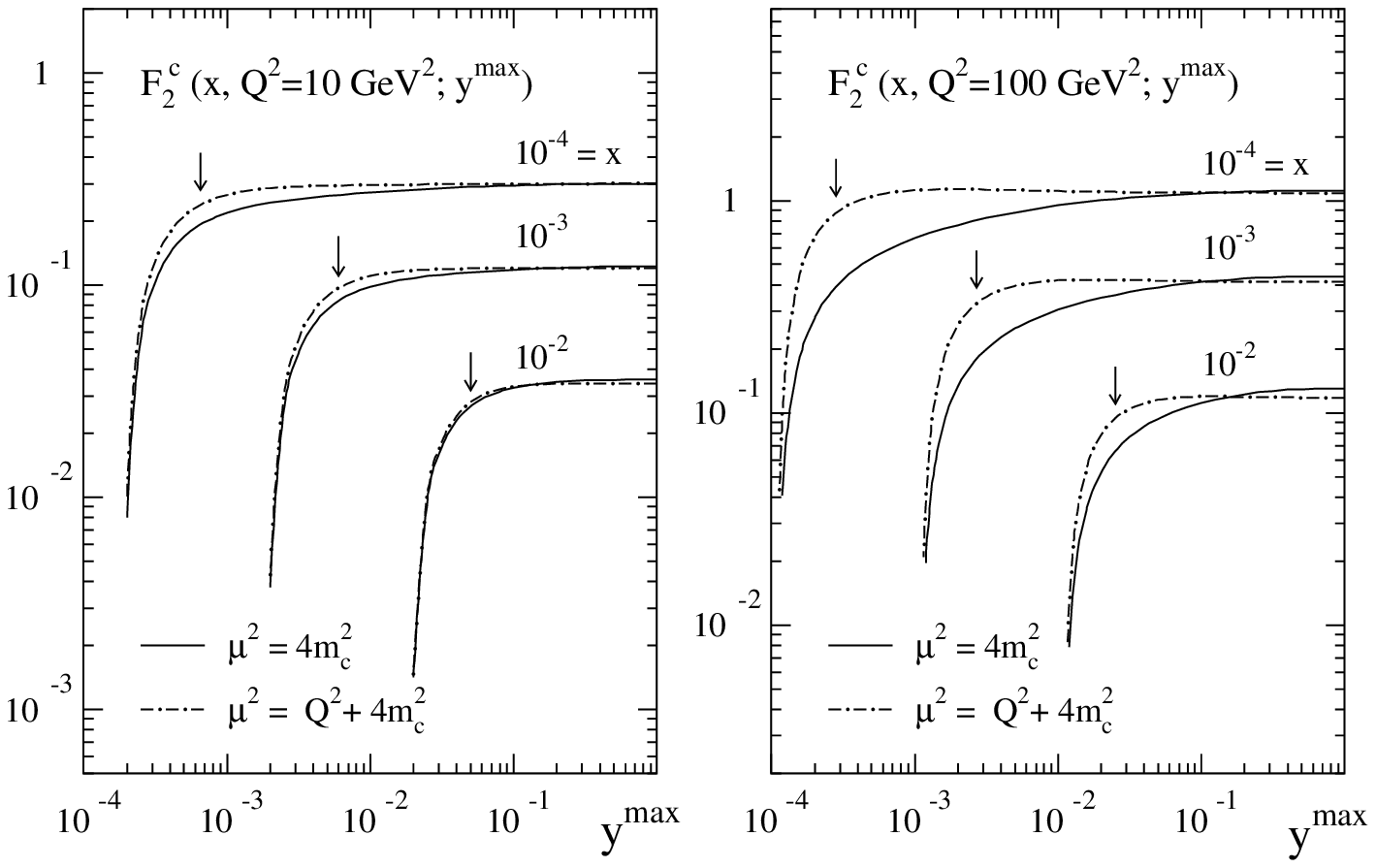,height=4.9cm,width=9.8cm}}
\vspace*{-4mm}
\end{center}
{\footnotesize \sf
  Figure 3: The contribution of the initial-parton momentum region
  $ ax \leq y \leq y^{\rm max} $ to $F_{2}^{\, c}$ at small $x$ for two
  choices of the scale $\mu $, using the parton densities of ref.~%
  \cite{GRV94}. The arrows indicate the values of $y^{\rm max}$ at
  which 80\% of the complete results are reached for $ \mu = \sqrt{Q^2
  + 4 m_{c}^{2}} $.}
\vspace{-3mm}
\end{figure}

\begin{figure}[htb]
\vspace{-5mm}
\begin{center}
\mbox{\epsfig{file=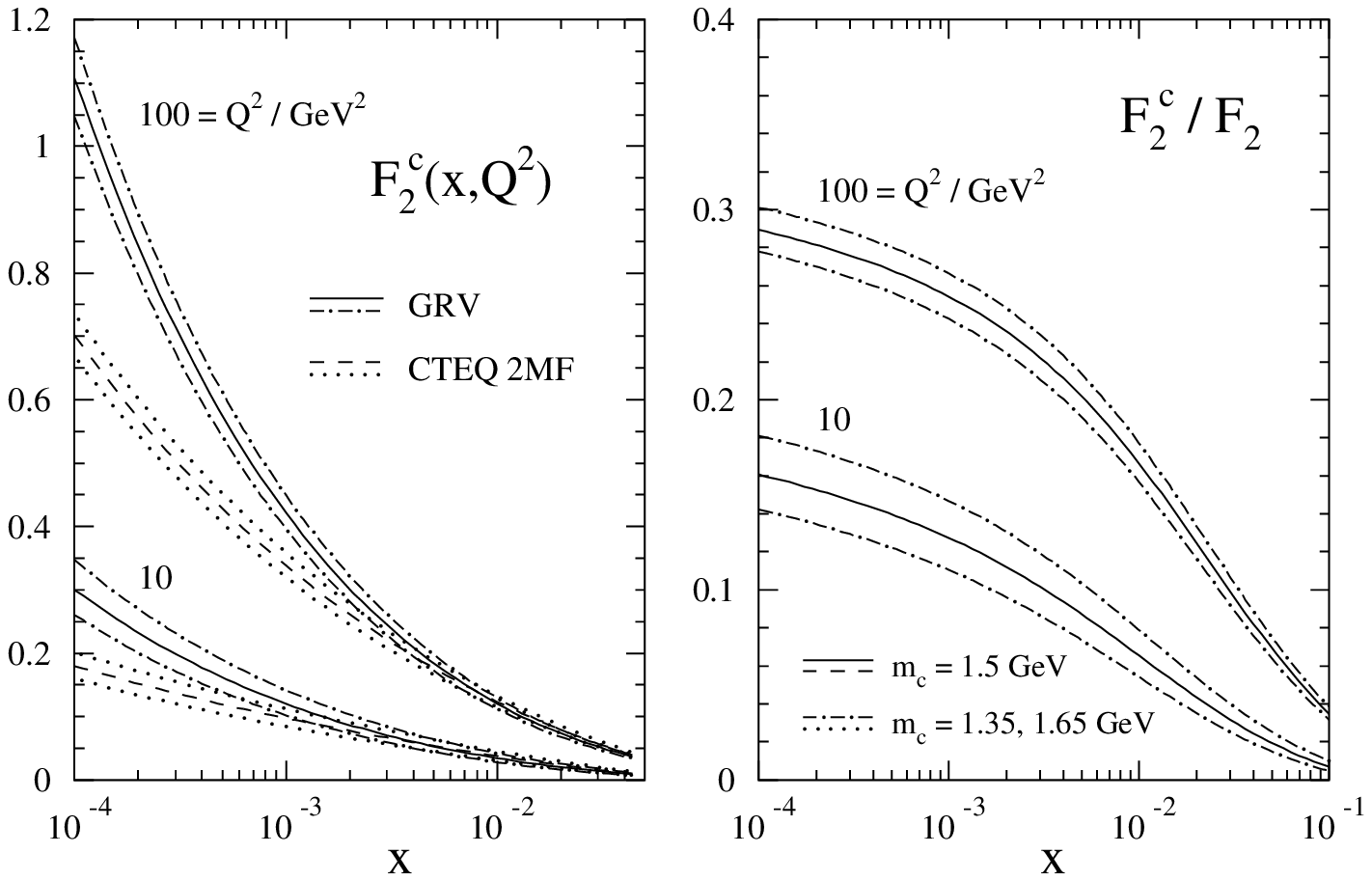,height=4.9cm,width=9.8cm}}
\vspace*{-4mm}
\end{center}
{\footnotesize \sf
  Figure 4: The $x$-dependence of $F_{2}^{\, c}$ and $ F_{2}^{\, c}/
  F_2 $ at two fixed values of $Q^2$, as expected from the GRV gluon 
  density \protect\cite{GRV94}. Also shown are $F_{2}^{\, c}$ as 
  obtained from the CTEQ$\, $2MF parton set \protect\cite{CTEQ} and the 
  charm mass dependence of the predictions. $ \mu = \sqrt{Q^2 + 4 m_{c}
  ^{2}} $ has been employed.}
\vspace{-4mm}
\end{figure}
The expected absolute and relative sizes of $F_{2}^{\, c}$ are displayed 
in Fig.~4. In contrast to the bottom contribution $F_{2}^{\, b}$ which 
reaches at most $ 2 \ldots 3\% $, $F_{2}^{\, c}$ is large in the HERA 
small-$x$ region, making up up to a quarter of $F_{2}$ as measured at 
HERA. See also ref.~\cite{LRSN1}. This size of $F_{2}^{\, c}$ renders a 
reliable (fully massive) treatment mandatory in any precise analysis of 
$F_2$ at small $x$.
The sensitivity of $F_{2}^{\, c}$ to the gluon density is illustrated by
the difference of the CTEQ$\, $2MF (flat $xg(x,\mu ^2 = 2.6 $ GeV$^2$) 
\cite{CTEQ} and the GRV (steep gluon) \cite{GRV94} expectations. There 
is quite some discriminative power of $F_{2}^{\, c}$, especially close 
to the lower end of the $Q^2$ range considered here, $Q^2 \approx 10 $ 
GeV$^2$. Moreover, by measuring up to about 100 GeV$^2$ in $Q^2$, the 
rapid growth of $yg(y \ll 1,\mu^2)$ predicted by the Altarelli-Parisi 
equations can be rather directly tested down to $ y \simeq 10^{-3} $.
A theoretical obstacle to an easy accurate gluon determination via the
charm structure function is the dependence of $F_{2}^{\, c}$ on the
unknown precise value of the charm quark mass $m_c$. A $\pm 10\%$ 
variation of $m_c$ also considered in Fig.~4 leads to a $ \pm 15 \ldots 
25\% \, (5 \ldots 10\%)$ effect at $Q^2 = 10 \, (100) $ GeV$^2$, 
respectively.

To summarize: 
The NLO perturbative QCD approach to the charm structure function $F_{2}
^{\, c}$ seems to be in good shape at small $x$ where $F_{2}^{\, c}$ is 
expected to be large (up to about a quarter of the total $F_{2}$ in the 
HERA regime), with scale variations of less than about $ \pm 10\% $. The 
situation is worse at large $x$, where the structure function is however 
small. 
$F_{2}^{\, c}$ represents a clean gluonic observable in NLO and is well 
suited for rather local a gluon measurement at small-$x$, with bulk of 
the result originating from gluon momenta within a factor of three above
the threshold value. Flavour excitation contributions from $\hat{s} \gg 
4m_{c}^{2} $ become important only at scales $Q^2$ higher than those 
relevant for small-$x$ observations at HERA considered here.
Despite the significant charm mass dependence of the results, especially
at low $Q^2$, useful constraints will be put on the proton's gluon 
density and its evolution at small $x$, if $F_{2}^{\, c}$ data with
an accuracy on the 10\% level can be obtained at HERA.

\vspace{1mm}
\noindent
{\bf Acknowledgement :} It is a pleasure to thank E. Laenen, S. Riemersma, 
J. Smith, and W. van Neerven for useful discussions. This work was 
supported by the German Federal Ministry for Research and Technology 
(BMBF) under contract No.\ 05 7WZ91P (0).

\end{document}